# IGGA: A Dataset of Industrial Guidelines and Policy Statements for Generative AIs

Junfeng Jiao [1], Saleh Afroogh*[2], Kevin Chen [3], David Atkinson [4], Amit Dhurandhar [5]

1. The University of Texas at Austin, Austin, TX 78712, United States. jjiao@austin.utexas.edu
2. The University of Texas at Austin, Austin, TX 78712, United States. Saleh.afroogh@utexas.edu
3. The University of Texas at Austin, Austin, TX 78712, USA. xc4646@utexas.edu
4. Allen Institute for AI (AI2), Seattle, USA davida@allenai.org
5. IBM Research Yorktown Heights, USA, adhuran@us.ibm.com

*Corresponding author(s): Saleh Afroogh (Saleh.afroogh@utexas.edu)

## Abstract

This paper introduces IGGA, a dataset of 160 industry guidelines and policy statements for the use of Generative AIs (GAIs) and Large Language Models (LLMs) in industry and workplace settings, collected from official company websites, and trustworthy news sources. The dataset contains 104,565 words and serves as a valuable resource for natural language processing tasks commonly applied in requirements engineering, such as model synthesis, abstraction identification, and document structure assessment. Additionally, IGGA can be further annotated to function as a benchmark for various tasks, including ambiguity detection, requirements categorization, and the identification of equivalent requirements. Our methodologically rigorous approach ensured a thorough examination, with a selection of reputable and influential companies that represent a diverse range of global institutions across six continents. The dataset captures perspectives from fourteen industry sectors, including technology, finance, and both public and private institutions, offering a broad spectrum of insights into the integration of GAIs and LLMs in industry.

## Background & Summary

Recent advancements in artificial intelligence (AI), particularly generative AI (GAI) technologies like Large Language Models (LLMs), have revolutionized industries worldwide, offering innovative tools for content creation, process optimization, and decision-making. However, the rapid adoption of GAI in professional contexts has also raised critical concerns about ethical and safe use, transparency, and the establishment of robust frameworks for AI governance. While many organizations are leveraging GAI technologies to enhance productivity and innovation, a gap remains in the development of comprehensive and standardized guidelines to manage their deployment responsibly A report by The Conference Board found that only 26% of organizations have a policy related to the use of generative AI, with an additional 23% developing such policies, underscoring the need for structured frameworks that balance technological advancements with ethical accountability.[1]

LLMs and GAI technologies are transforming workplace dynamics by enabling the generation of sophisticated, human-like content through advanced natural language processing (NLP) techniques. These models are powered by neural network architectures that learn from extensive datasets and can generate contextually relevant and semantically coherent content. By leveraging vast datasets and billions of parameters, LLMs have made it possible to predict linguistic patterns, enabling applications such as automated reporting, virtual assistance, and content strategy. However, the lack of standardized policies governing these tools across industries poses questions in ensuring consistent and responsible implementation.

To address these challenges, **IGGA (IGGA: A Dataset of Industrial Guidelines and Policy Statements for Generative AIs)** introduces a curated dataset of 160 guidelines and policy statements from leading companies spanning 14 industrial sectors across 7 continents. This dataset consolidates AI-related policies from globally recognized organizations, providing a unique resource for analysing and developing best practices in AI governance, workplace integration, and management. IGGA's scope covers diverse domains, including technology, healthcare, finance, manufacturing, and education, ensuring comprehensive coverage of industry-specific approaches to AI.

To validate the quality and applicability of the IGGA dataset, two qualitative analyses were conducted. The first analysis examines the dataset's geographic and sectoral diversity, ensuring representation across continents and industries. The second focuses on text-based analyses, evaluating the structure, clarity, and content of the guidelines. By leveraging NLP techniques, these analyses highlight the dataset's ability to support tasks such as policy synthesis, requirement categorization, ambiguity detection, and benchmarking.

The IGGA dataset aims to facilitate replicable and scalable NLP experiments while promoting the creation of generalizable insights across industries. To enhance its utility, recommendations for future expansions are provided, including the addition of emerging industry guidelines and updates to existing documents. These expansions will be made available on GitHub, enabling broader accessibility and collaboration among researchers, policymakers, and industry professionals.

**Figure 1.** Schematic Overview of the IGGA Study Design: Data Collection, Structuring, Analysis, and Utilization

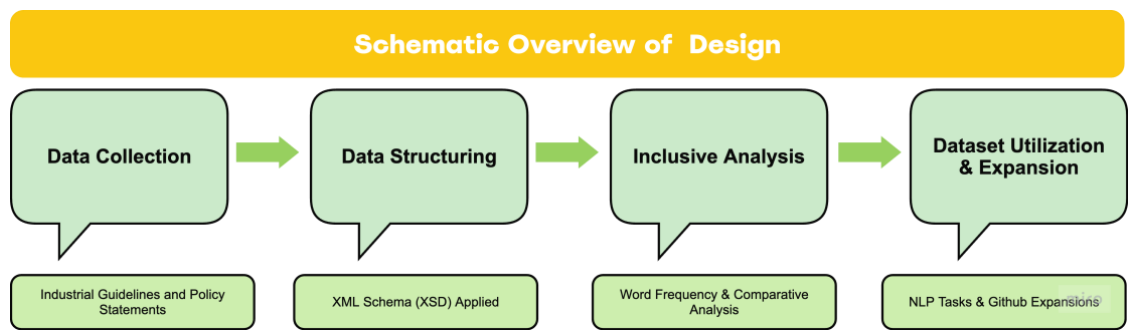

**Figure 1**. provides a clear and aesthetically pleasing representation of the key stages in your research:

1. Data Collection: Gathering guidelines from 160 universities across 14 different industries in 7 continents
2. Data Structuring: Applying an XML Schema (XSD) to standardize the documents.
3. Inclusive Analysis: Conducting word frequency analysis and assessing document structure.
4. Dataset Utilization & Expansion: Using the dataset for NLP tasks and planning to extend it by adding new guidelines on GitHub.

## Methods

In this study, we conducted a comprehensive data collection of industry guidelines and policy statements for the use of Generative AIs (GAIs) and Large Language Models (LLMs) in workplace and management contexts. Our data collection phase focused on gathering official guidelines and policy statements from 160 leading companies worldwide, ensuring a representative sample across various industrial sectors and geographical regions. The selection criteria for companies included industry leadership, geographical diversity, sectoral representation, and the presence of official guidelines and policy statements specifically addressing the use of GAIs, with an emphasis on workplace integration and management strategies. Companies lacking such guidelines were excluded, and replacements were selected based on similar criteria to maintain consistency and reliability in our analysis. This rigorous selection process was instrumental in ensuring the credibility and applicability of our findings (See, Table 1).

**Table 1.** Analytics Workflow of the analysis of GAI/LLM guidelines and policy statements

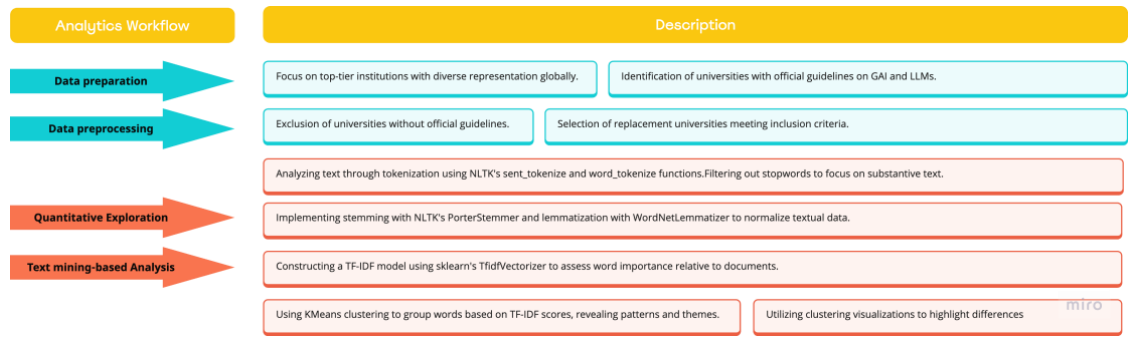

### Categorization Criteria

For finding industry-specific recommendations, we utilized two primary categorization criteria: geographical location and industrial sector. Geographical location explores the influence of regional factors, regulatory environments, and cultural contexts on guideline and policy statements development and application. We ensured diversity within each continent, recognizing the impact of cultural backgrounds and regional regulatory frameworks on the formulation and adoption of guidelines and policy statements. Additionally, we included guidelines from a variety of sectors—such as technology, healthcare, manufacturing, finance, and retail—providing a comprehensive and cross-sectoral view of GAI integration.

Industrial sector categorization was particularly significant in understanding the practical aspects, challenges, and sector-specific considerations for guideline implementation. This dual categorization approach enabled a holistic exploration of GAI/LLM guidelines and policy statements, encompassing specialized insights from diverse industries alongside broader global perspectives that influence AI governance and management practices.

Supplementary Table 1 presents 160 official guidelines and policy statements collected from companies across 14 industrial sectors and seven continents, reflecting a wide array of cultural, industrial, and geographical perspectives. By highlighting the interplay between regional influences and sectoral needs, this dataset offers a robust foundation for analysing the varied approaches to GAI and LLM adoption in professional settings.

### Text Mining and Computational Processing

To further analyze the collected guidelines and policy statements, we employed a text mining-based approach. The analysis began with tokenization, where the text data from the 160 guidelines and policy statements was broken down into sentences and subsequently into individual words. This process, facilitated by the Natural Language Toolkit (nltk), involved using sent_tokenize and word_tokenize functions, enabling us to scrutinize the text at a granular level.

To refine the data further, we filtered out stopwords using nltk's stopwords list, thereby eliminating commonly used words that add little semantic value and focusing on substantive text..After tokenization, stemming and lemmatization were used to to normalize the textual data. Stemming, which is performed using nltk's PorterStemmer, reduced words to their root form, grouping different forms of the same word. Lemmatization, using WordNetLemmatizer, converted words into their dictionary form, for context-aware normalization of the text. These methods were usedin enhancing the accuracy of frequency counts during the analysis.

With a cleansed and normalized dataset, we constructed a Term Frequency-Inverse Document Frequency (TF-IDF) model using sklearn's TfidfVectorizer. This model balanced the frequency of words (Term Frequency) against the rate of their appearance across multiple documents (Inverse Document Frequency), allowing us to assess the significance of words within specific documents. The TF-IDF model was pivotal in highlighting key terms within the academic guidelines and policy statements.

Subsequently, we utilized the KMeans clustering algorithm to group words based on their TF-IDF scores, revealing patterns and themes within the text. Clustering visualizations illustrated the prominence of different terms, providing insights into the commonalities and distinctions across the guidelines and policy statements.

### Data Exclusion and Ethical Considerations

Companies without official guidelines and policy statements on their websites or from relevant sources that the specifically addressing the usage of GAIs and LLMs in the workplace and management were excluded from this study. These exclusion criteria were established prior to the study to ensure the methodological consistency and reliability of our findings. Replacement companies were carefully selected to maintain the representativeness of the sample across industries and regions.

This study did not involve human or animal subjects, and therefore ethical approvals were not required. The data collected and analyzed were publicly available from official company websites, ensuring that no sensitive or proprietary information was utilized in this research. This approach guarantees transparency and adherence to ethical research practices.

### Computational Tools

All processing of text data and visualizations was conducted using Python 3.8 and Jupyer notebook v7.2, with key libraries including nltk for natural language processing and sklearn for machine learning tasks. The text mining and clustering analyses were performed on a high-performance computing environment to ensure in handling the large dataset.o

### Data Records

The IGGA dataset, titled " IGGA: A Dataset of Industrial Guidelines and Policy Statements for Generative AIs," is available on the Harvard Dataverse repository. It is associated with the Persistent Identifier doi:10.7910/DVN/4LOXUW and was published on June 4, 2024. The dataset is available in its fourth version, contributed by scholars the University of Texas at Austin, the Allen Institute for AI, and IBM Research. The dataset is licensed under the Creative Commons CC0 1.0 Universal Public Domain Dedication (CC0 1.0), allowing for unrestricted use and distribution, provided proper credit is given.

### Overview of Data Files

The IGGA dataset is available in three file formats: MS Word, PDF, and MS Excel, each tailored to different analytical requirements. The MS Word document, titled "IGGA: A Dataset of Industrial Guidelines and Policy Statements for Generative AIs.docx," is a 1 MB file published on November 19, 2024. It comprises three sections: a table of 160 academic guidelines and policy statements, citations for each guideline and policy statement, and the full text of all 160 guidelines, organized with a comprehensive table of contents. Similarly, the PDF file, "IGGA: A Dataset of Industrial Guidelines and Policy Statements for Generative AIs.pdf," replicates the structure of the Word document. This 3.9 MB file, also published on November 19, 2024, is designed for convenient access and reference. The Excel file, "IGGA.xlsx," published on November 17, 2024, is a compact 33.8 KB spreadsheet that offers a structured dataset with fields such as Guideline ID, Continent, Country, University, Document/Website Name, and Number of Pages. These formats address diverse user needs, supporting both in-depth textual analysis and organized data extraction

### File Structure and Format Details

The IGGA dataset files are available in MS Word and PDF files provide a comprehensive textual representation of the industry guidelines and policy statement, with each document systematically sectioned for ease of navigation. These files include a detailed table of contents, facilitating straightforward access to specific sections. The content is divided into three primary sections: a table summarizing the industry guidelines, citations for each guideline and policy statement, and the full text of the guidelines and policy statements. This structure supports both in-depth qualitative analysis and quick reference for practical applications.

The Excel file offers a detailed, categorized view of the dataset, organizing guidelines and policy statements by industrial sectors and geographical regions. This format is particularly valuable for quantitative analysis and enables cross-referencing with other datasets. Its structured layout allows users to sort and filter data easily, enabling tailored analyses based on specific research or organizational needs. The combination of these formats ensures that the IGGA dataset is highly versatile and accessible for a variety of research and industry-oriented purposes.

### Data Citation

To ensure proper attribution and to adhere to academic standards, each data file associated with the IGGA dataset should be cited as follows: " Jiao, Junfeng; Afroogh, Saleh; Chen, Kevin; Atkinson, David; Dhurandhar, Amit, 2024, "IGGA: A Dataset of Industrial Guidelines and Policy Statements for Generative AIs", https://doi.org/10.7910/DVN/4LOXUW, Harvard Dataverse, V2." [2]

### Technical Validation

To validate the technical quality and inclusiveness of the IGGA dataset, two qualitative analyses were conducted as part of the dataset's comprehensive description phase. These analyses provide a descriptive overview of the dataset's capacity to support NLP-based analyses of industry guidelines and policy statements. The first analysis evaluates the inclusiveness of the dataset, showcasing its extensive coverage across seven continents and 14 industrial sectors.

The second analysis involves a text-based exploration and visualization of the guidelines, categorized by industrial sectors and types of recommendations, which will be discussed in greater detail in subsequent sections.

### Text-Based Analysis and Visualization Across Fourteen Industries

The IGGA dataset includes industry guidelines and policy statements from fourteen sectors across seven continents: Africa, Asia, Europe, North America, South America, Oceania, and Antarctica (via multinational industry representation). Textual data for these guidelines and policy statements was extracted from the primary IGGA dataset document, available in DOCX format. The preprocessing steps were executed using Python, with the necessary libraries installed in Jupyter Notebook.

The initial step involved converting the DOCX files into plain text using the docx2txt library. This conversion ensured standardization of the text format, making it easily readable and ready for subsequent processing. The extracted text content was then subjected to several preprocessing steps, including tokenization, cleaning, and formatting, to prepare it for qualitative and quantitative analyses. These preprocessing methods ensure the dataset is reliable, clean, and ready for NLP-based exploration and application.

### Text Processing

The IGGA dataset underwent several preprocessing steps to prepare the text data for analysis. The text was initially read from converted TXT files and transformed to lowercase to ensure consistency. Non-alphanumeric characters and numbers were removed to clean the data and eliminate irrelevant elements. This was followed by tokenization, where the text was divided into sentences and further into individual words or phrases for analysis. Stopwords, such as "and," "the," and "is," were removed to focus on analytically relevant terms, along with additional industry-specific terms that were deemed non-essential.

Next, stemming was applied using the Porter Stemmer to reduce words to their root forms, grouping variations of the same word under a single base form. Lemmatization was performed using the WordNetLemmatizer to refine these root forms into their most basic and valid structure while retaining grammatical accuracy. This combination of stemming and lemmatization ensured uniformity in the text, allowing for more consistent analysis across the dataset.

The processed text was stored in structured files for further analysis. These steps resulted in a standardized and clean dataset, ready for tasks such as topic modeling, sentiment analysis, and keyword extraction. This preprocessing pipeline provides a foundational framework for analyzing industry-specific guidelines and policy statements for AI management and workplace integration across sectors and regions.

### Frequency Analysis and Keyword Extraction

Following preprocessing, we identified the twenty most popular keywords for each of the 14 indisutry sectors . This step involved tallying the occurrences of each word after stopwords were removed and selecting the top 20 words with the highest frequencies. These keywords were then visualized using frequency charts, which graphically displayed the most frequently occurring words for each industry to assess the variety of different words each sector utilizes in their guidelines and policy statements.

**Figure 1**: The Frequency of Key Concepts in six continents

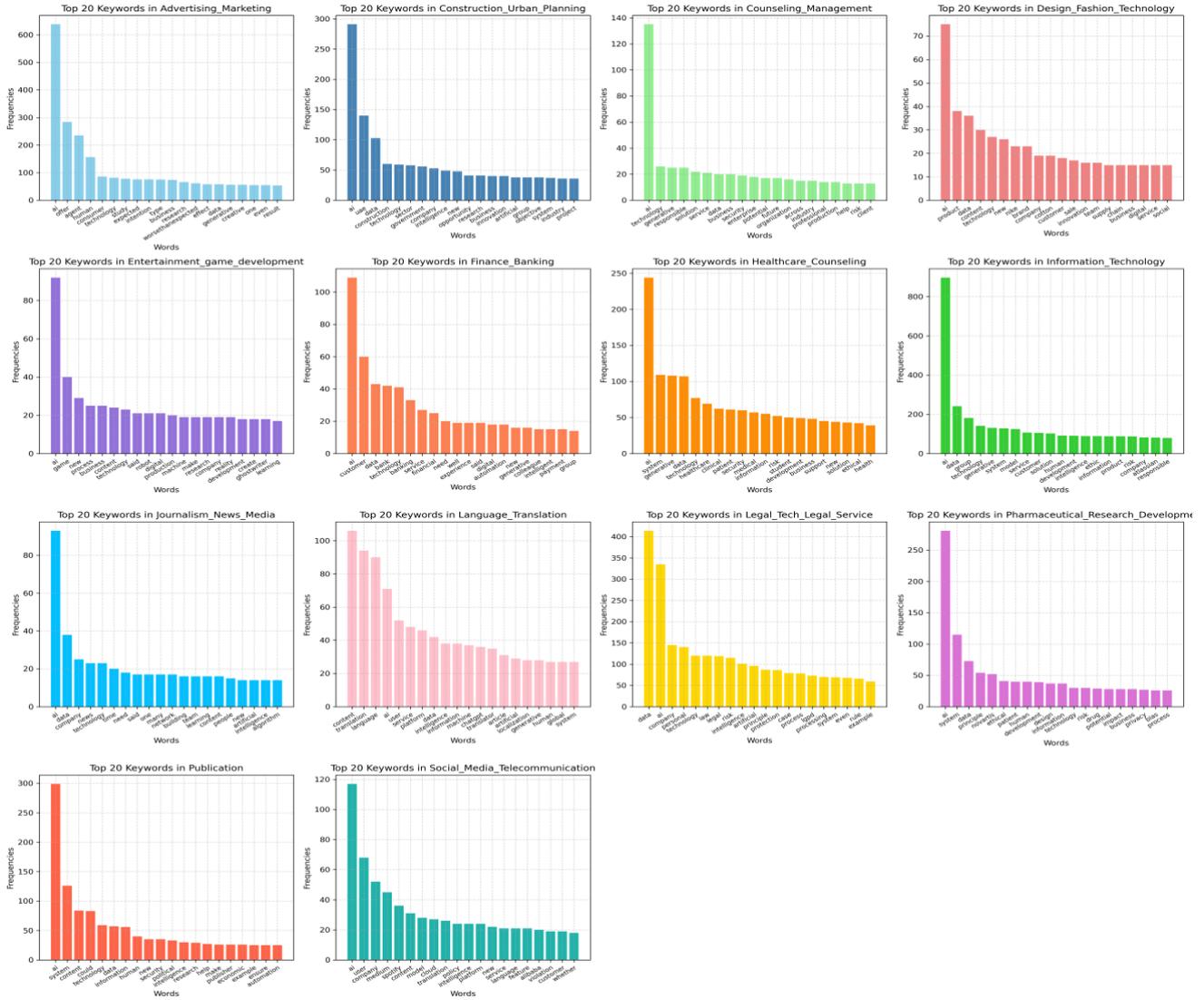

## Term Frequency-Inverse Document Frequency (TF-IDF) Heatmap Analysis

We conducted a quantitative semantic analysis using TF-IDF and cosine similarity to quantify and visualize the thematic overlaps across industries. By converting guideline and policy statements documents into vector representations, it identifies terms that are both commonly emphasized and unique to specific industries. The cosine similarity matrix, depicted as a heatmap, provides an intuitive comparison of the guideline and policy statement alignment between industries, highlighting areas of shared focus and divergence. Figure two illustrates the heatmap similarity analysis of these quantitative findings, offering insights into commonalities and differences in AI priorities across sectors.

**Figure 2: Heatmap Analysis Similarity Scores across Fourteen Industries**

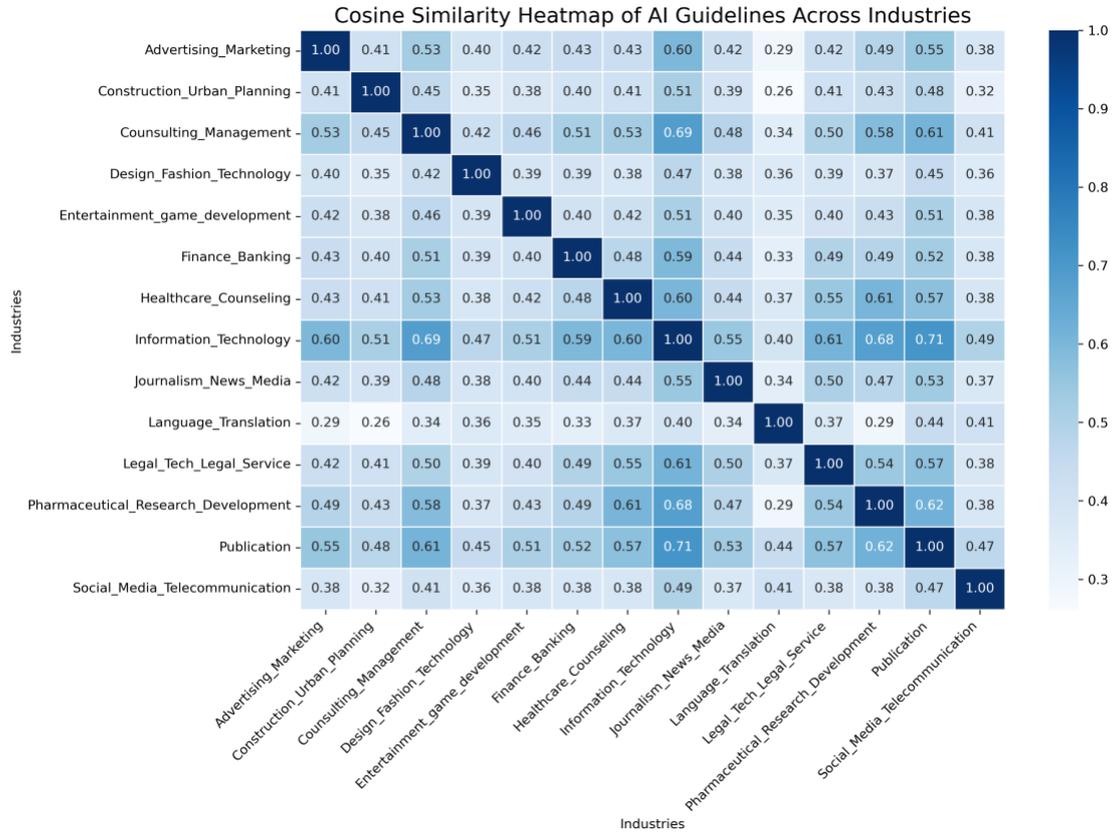

**Validation of Dataset Inclusiveness and Global Representation**

The cosine similarity heatmap effectively illustrates the thematic overlaps and distinctions in AI guidelines and policy statements across 14 industrial sectors, highlighting shared and sector-specific priorities. For example, sectors such as Information Technology, Publication, and Pharmaceutical Research & Development exhibit significant thematic alignment, with similarity scores ranging from 0.60 to 0.71, reflecting common global themes like data analytics, technological integration, and innovation.

In Advertising & Marketing, the analysis reveals a focus on leveraging AI for customer engagement and personalized content, reflected in moderate similarity scores with Information Technology (0.60) and Publication (0.55). In Construction & Urban Planning, themes like "automation" and "infrastructure optimization" emerge, but low similarity scores with other sectors (e.g., 0.26 with Language Translation) reflect its unique operational focus.

The Consulting & Management sector highlights AI's role in decision support and operational efficiency, achieving high similarity scores with Information Technology (0.69) and Healthcare

& Counseling (0.53), emphasizing shared concerns about ethical and safe AI implementation and privacy. Entertainment & Game Development shows moderate similarity with Information Technology (0.51) and Publication (0.52), reflecting the role of AI in creative production and user engagement.

Finance & Banking prioritizes AI for risk management and predictive analytics, achieving notable similarity scores with Healthcare & Counseling (0.48) and Pharmaceutical Research & Development (0.49), underlining shared themes of data security and compliance. Legal & Tech Services, with a focus on governance and regulatory frameworks, shows a moderate overlap with Pharmaceutical Research & Development (0.49) and Publication (0.57).

The Language Translation sector stands out with low similarity scores across most industries, such as 0.26 with Construction & Urban Planning, emphasizing its unique focus on language-specific applications of AI. Similarly, Social Media & Telecommunication exhibits limited thematic overlap with sectors like Construction & Urban Planning (0.32) but aligns better with Information Technology (0.49) due to shared reliance on AI for data handling and communication technologies.

Overall, this heatmap-based analysis demonstrates the IGGA dataset's inclusiveness and potential to support NLP-based exploration of industry-specific AI guidelines and policy statements. The diversity of themes across sectors highlights the dataset's capacity to provide insights into shared global priorities, such as ethics, safety, and innovation, while capturing unique sectoral needs, such as compliance in legal services or creative AI applications in entertainment. This balance of shared and distinct themes validates the dataset's utility for analyzing AI adoption and governance in diverse industrial contexts.

## Usage Notes

The IGGA dataset is a versatile resource for researchers and practitioners exploring industry guidelines and policy statements for the use of Generative AIs (GAIs) and Large Language Models (LLMs) in workplace and management settings. To maximize the utility of this dataset, we provide several recommendations and tips for effective use.

Researchers can leverage a variety of software tools and libraries to analyze the dataset. Python is particularly well-suited for text-based analyses, offering comprehensive libraries such as nltk for natural language processing tasks, sklearn for machine learning applications, , sneaky and pyplot from matplotlib for data visualization and plotly for interactivity. The dataset's formats—DOCX, PDF, and Excel—ensure compatibility with standard text processing workflows and spreadsheet software, enabling seamless import for analysis.

For those interested in deeper analyses, we suggest several advanced techniques. Text normalization can help standardize the data for consistent comparison across guidelines and policy statements. Additionally, methods such as topic modeling, clustering, or sentiment analysis can be employed to uncover prevalent themes, patterns, and perspectives within the guidelines and policy statements. These approaches can provide valuable insights into industry-specific trends and global attitudes toward GAI and LLM implementation.

## Privacy and Safety Considerations

The IGGA dataset comprises publicly available industry guidelines and policy statements collected from official company websites. As such, there are no privacy or safety concerns associated with accessing this data. Researchers are encouraged to use the dataset responsibly, adhering to ethical and safety standards relevant to their fields. The dataset is freely available under the Creative Commons CC0 1.0 Universal Public Domain Dedication, ensuring unrestricted public access.

## Code Availability

The IGGA dataset was processed and analyzed using custom Python scripts within a Jupyter Notebook environment. These scripts include tools for converting DOCX files to plain text, text preprocessing, and performing various text-mining analyses. Researchers interested in accessing the code can contact the corresponding author, Saleh Afroogh, through the provided contact details.

The scripts were developed using Python 3.8 and rely on several key libraries, including nltk (version 3.5) for natural language processing, sklearn (version 0.24) for machine learning, and matplotlib (version 3.3) for data visualization. Documentation within the code comments clarifies the specific parameters and variables used, such as stopword lists, stemming rules, and lemmatization techniques, ensuring reproducibility and adaptability.

To support transparency and facilitate future research, the scripts and any updates or modifications will be shared through the authors' GitHub repository, with proper version control to track changes. This ensures that the tools remain accessible and useful for researchers looking to replicate or extend the analysis.

## Acknowledgements

This research was funded by the National Science Foundation under grant number 2125858. The authors would like to express their gratitude for the foundation's support, which made this study possible. Furthermore, in accordance with MLA, we would like to thank OpenAI for its assistance in editing.

## Competing interests

The authors declare that the research was conducted in the absence of any commercial or financial relationships that could be construed as a potential conflict of interest.

**Supplementary Table 1**: Industry-level guidelines and policy statements for Usage of LLMs and GAIs

| | Industry | | Company | Name of document/website | Country | Continent |
|---|---|---|---|---|---|---|
| 1 | Healthcare Counseling (10) | 1 | Becker's Healthcare | Should health systems regulate the use of ChatGPT?[3] | USA | North America |
| | | 2 | Orbita Hospitals and Health Care | OpenAI and ChatGPT: A Primer for Healthcare Leaders[4] | USA | |
| | | 3 | Chugai Pharmaceutical Co | Chugai DX Meeting[5] | Japan | Asia |
| | | 4 | Daiichi Sankyo Company | Daiichi Sankyo's Challenge to Realize: 2030 Version[6] | Japan | |
| | | 5 | Hardian Health | How to get ChatGPT regulatory approved as a medical device[7] | UK | Europe |
| | | 6 | AstraZeneca | AstraZeneca data and AI ethics[8] | UK | |
| | | 7 | Procaps Group | Procaps participate in panels on digitalization and artificial intelligence in pharma manufacturing[9] | Colombia | South America |
| | | 8 | 1DOC3 | 1DOC3: ACCESSIBLE HEALTHCARE TO MILLIONS[10] | Colombia | |
| | | 9 | CSL Limited | Artificial Intelligence at CSL[11] | Australia | Australia |
| | | 10 | DokiLink | Artificial Intelligence in Africa's Healthcare: Ethical Considerations[12] | South Africa | Africa |
| 2 | Technology and IT Services (30) | 11 | Microsoft | Artificial Intelligence (AI) usage policy for Microsoft[13] | USA | North America |
| | | 12 | Apple | Apple Restricts Use of ChatGPT[14] | USA | |
| | | 13 | Alphabet | Google AI principles[15] | USA | |
| | | 14 | Meta Platforms, Inc. | Meta Guideline to Responsible AI[16] | USA | |
| | | 15 | Amazon | AWS Responsible AI Policy[17] | USA | |
| | | 16 | Oracle | Oracle Generative AI strategy[18] | USA | |
| | | 17 | Adobe | Adobe Generative AI Guideliens[19] | USA | |
| | | 18 | Qualcomm | Qualcomm AI strategy[20] | USA | |
| | | 19 | Salesforce | Generative AI: 5 Guidelines for Responsible Development[21] | USA | |
| | | 20 | CGI Inc | CGI optimizing generative AI potential[22] | Canada | |
| | | 21 | LG Electronics | LG Presents 'AI Ethics Principles' for Trustworthy AI Research[23] | South Korea | Asia |
| | | 22 | Fujitsu | Fujitsu Generative AI use guidelines[24] | Japan | |
| | | 23 | Panasonic | Panasonic Responsible AI use[25] | Japan | |
| | | 24 | Sony | Sony Responsible AI Usage[26] | Japan | |
| | | 25 | Asm Pacific Technology | ASMPT 2023 Interim Report[27] | Hong Kong | |
| | | 26 | Infosys | Infosys Responsible AI[28] | India | |
| | | 27 | Samsung Group | Samsung Group Digital Responsibility[29] | South Korea | |
| | | 28 | Sea Limited (Garena) | Sea Limited – Digital Transformation Strategies[30] | Singapore | |
| | | 29 | Grab | Grab's AI Ethics Principles[31] | Singapore | |
| | | 30 | Sea Group | Sea Founder Warns of Turmoil From the Shift to AI[32] | Singapore | |
| | | 31 | Atos | Atos blueprint for generative AI[33] | France | Europe |
| | | 32 | Capgemini | AI code of ethics.[34] | France | |
| | | 33 | Dormakaba | AI based anti-tailgating Solution[35] | Switzerland | |
| | | 34 | Ericsson | trustworthy AI[36] | Sweden | |
| | | 35 | Accenture | AI ethics & governance[37] | Ireland | |
| | | 36 | SAP | SAP AI ethics[38] | Germany | |
| | | 37 | amadeus it | Capturing the power of Generative Artificial Intelligence to enhance the passenger experience[39] | Spain | South America |
| | | 38 | Gorilla Logic | Ai in product development[40] | Costa Rica | |
| | | 39 | Atlassian | Atlassian Intelligence is built on trust[41] | Australia | Australia |
| | | 40 | Flutterwave | Flutterwave strategic agreement[42] | Nigeria | Africa |
| 3 | Finance and Banking (10) | 41 | JPMorgan Chase | JPMorgan Chase Restricts Staffers' Use Of ChatGPT[43] | USA | North America |
| | | 42 | Wells Fargo | Wells Fargo, artificial intelligence, and you[44] | USA | |
| | | 43 | Mizuho Financial Group | Mizuho permits 45,000 employees to use generative AI[45] | Japan | Asia |
| | | 44 | State Bank of India | SBI Embraces AI and ML Technologies to Transform Banking Operations[46] | Japan | |
| | | 45 | HSBC Holdings Plc | HSBC principles for ethical use of data and AI[47] | UK | Europe |
| | | 46 | Lloyds Banking Group | How AI automation is helping[48] | UK | |
| | | 47 | Itaú Unibanco Holding | Brazil's Itaú in 'very good position' to harness generative AI[49] | Brazil | South America |
| | | 48 | Banco de la Nacion Argentina | Ethical and responsable use of AI in Argentina worker's rights[50] | Argentina | |
| | | 49 | Westpac | How AI will shape the future of banking[51] | Australia | Australia |
| | | 50 | Standard Bank Group | The Future of Digital Banking and Transacting in Africa[52] | South Africa | Africa |
| 4 | Publication Industry (10) | 51 | Elsevier | The use of generative AI and AI-assisted technologies in writing for Elsevier[53] | USA | North America |
| | | 52 | The New York Times Company | AI workplace changes[54] | USA | |
| | | 53 | China Daily | Guidelines establish proper uses of AI in research[55] | China | Asia |
| | | 54 | South China Morning Post | China unveils new artificial intelligence guidelines [56] | China | |
| | | 55 | Penguin Random House | Penguin Random House CEO hopes AI will help sell more books: Report[57] | UK | Europe |
| | | 56 | Wolters Kluwer | Artificial Intelligence (AI) Principles[58] | Netherland | |
| | | 57 | Grupo Planeta | AI policy[59] | Spain | South America |
| | | 58 | Grupo Editorial Record | Principles for AI[60] | Brazil | |
| | | 59 | Allen & Unwin | AI and International security[61] | Australia | Australia |
| | | 60 | News24 | News24 to use AI in moderating comments[62] | South Africa | Africa |
| 5 | Language Translation Services (10) | 61 | American Translators Association | ChatGPT for Translators: How to Use the Tool to Work More Efficiently?[63] | USA | North America |
| | | 62 | TransPerfect | TransPerfect generative AI[64] | USA | |
| | | 63 | SEAtongue | Using AI for interpretation[65] | Malaysia | Asia |
| | | 64 | Jinyu Translation | Translation: Measures for the Management of Generative Artificial Intelligence[66] | China | |
| | | 65 | SDL plc | SDL to expand knowledge discovery and intelligent process automation to additional languages[67] | UK | Europe |
| | | 66 | Lionbridge | Lionbridge usage of AI[68] | Ireland | |
| | | 67 | Altura Interactive | ChatGPT translation[69] | Mexico | South America |
| | | 68 | GLOBO Brazil | Globo Pacts with Google Cloud in Bid to Become a Mediatech Company[70] | Brazil | |
| | | 69 | Straker Translations | Traker translations security[71] | New Zealand | Australia |
| | | 70 | Elite Translations Africa | Harness the power to AI for preservation of African languages[72] | Kenya | Africa |
| 6 | Construction and Urban Planning (10) | 71 | Bechtel | Applications of Artificial Intelligence In EPC[73] | USA | North America |
| | | 72 | Turner Construction Company | AI at 2023 summit[74] | USA | |
| | | 73 | Hyundai E&C | Hyundai E&C ensures Safety and Quality Management of construction sites[75] | South Korea | Asia |
| | | 74 | Shimizu Corporation | Shimizu develops AI systems for initial structural designs[76] | Japan | |
| | | 75 | Vinci SA | Innovation and prospective[77] | France | Europe |
| | | 76 | Bouygues Construction | Innovation strategy[78] | France | |
| | | 77 | Grupo ACS | Shareholder's meeting & future prospects[79] | Spain | South America |
| | | 78 | Sacyr | IntegratedSustainabilityReport 2023 – Committed to a Sustainable Future[80] | Argentina | |
| | | 79 | Lendlease | Lendlease CEO on using AI in their businesses[81] | Australia | Australia |
| | | 80 | Arab Contractors | UAE National strategy for generative AI[82] | Egypt | Africa |
| 7 | Consulting and management (10) | 81 | McKinsey | About half of McKinsey staff allowed to use generative AI: report[83] | USA | North America |
| | | 82 | Deloitte | Deloitte Launches Innovative 'DARTbot' Internal Chatbot[84] | USA | |
| | | 83 | Tata Consultancy Services | The future is AI. The future is human[85] | India | Asia |
| | | 84 | Infosys | Responsible ai[86] | India | |
| | | 85 | Roland Berger | ChatGPT a game changer for artificial intelligence[87] | Germany | Europe |
| | | 86 | Grupo Assa | Digital transformation practice[88] | Brazil | South America |
| | | 87 | Falconi Consultores de Resultado | Optimizing performance with artificial intlleigence[89] | Brazil | |
| | | 88 | Nous Group | AI powering innovation and productivity [90] | Australia | Australia |
| | | 89 | Africa International Advisors | Future of Africa International Advisors Group[91] | South Africa | Africa |
| 8 | Design and Fashion Technology (10) | 90 | Nike | Nike leveraging AI in operations[92] | USA | North America |
| | | 91 | Ralph Lauren Corporation | Ralph Lauren testing AI[93] | USA | |
| | | 92 | Comme des Garçons | AI designer creating fashion grails from iconic runways[94] | Japan | Asia |
| | | 93 | Shiseido Company, Limited | AI and change in management[95] | Japan | |
| | | 94 | LVMH Moët Hennessy Louis Vuitton SE | Joins stanford developement program[96] | France | Europe |
| | | 95 | Kering | AI and innovation[97] | France | |
| | | 96 | Havaianas | AI powered Case study[98] | Brazil | South America |
| | | 97 | Fatabella | Falabella hires Amelia as a digital assistant for its employees[99] | Chile | |
| | | 98 | Cotton On Group | How Cotton On Is Taking the Aussie Aesthetic Global with AI[100] | Australia | Australia |
| | | 99 | David Tlale | South Africa to adopt AI[101] | South Africa | Africa |
| 9 | Entertainment and game development (10) | 100 | Activision Blizzard | AI in game development[102] | USA | North America |
| | | 101 | Electronic Arts | AI in game industry[103] | USA | |

| | | 102 | Tencent | Tencent AI policy[104] | China | Asia |
|---|---|---|---|---|---|---|
| | | 103 | Square Enix | Letter from the president on AI[105] | Japan | |
| | | 104 | Ubisoft | Ghostwriter using AI in script writing[106] | France | Europe |
| | | 105 | Supercell | Supercell AI fund venture[107] | Finland | |
| | | 106 | Wildlife Studios | AI trust and safety[108] | Brazil | South America |
| | | 107 | Globant | Globant AI Manifesto[109] | Argentina | |
| | | 108 | Wargaming Sydney | AI in Wargaming[110] | Australia | Australia |
| | | 109 | Kukua | Future production with AI[111] | Kenya | Africa |
| 10 | Journalism and News Media (10) | 110 | New York Times | The New York Times is building a team to explore AI in the newsroom[112] | USA | North America |
| | | 111 | ABC News | ABC builds its own AI model[113] | USA | |
| | | 112 | JoongAng Daily | JoongAng Group Builds South Korea's First AI-Driven Enterprise Network by Juniper[114] | South Korea | Asia |
| | | 113 | Asahi Shimbun Company | Panel Discussion Artificial Intelligence and Democracy[115] | Japan | |
| | | 114 | BBC | BBC AI Principles[116] | UK | Europe |
| | | 115 | Reuters | Reuter AI ethics and principles[117] | UK | |
| | | 116 | Grupo Globo | Grupo Globo CEO on Evolving rules and regulations surrounding AI[118] | Brazil | South America |
| | | 117 | Pharu | Pharu and his challenge of bringing analytics to Latin American culture[119] | Chile | |
| | | 118 | News Corp Australia | News Corp AI powered News[120] | Australia | Australia |
| | | 119 | Nation Media Group | State in battle to protect data privacy, enhance security in the AI age[121] | Kenya | Africa |
| 11 | Pharmaceutical Research and Development (10) | 120 | Pfizer Inc. | Pfizer AI policy and position[122] | USA | North America |
| | | 121 | Johnson & Johnson | Jnj policy and positions[123] | USA | |
| | | 122 | Takeda Pharmaceutical Company Limited | Takeda position on AI[124] | Japan | Asia |
| | | 123 | Eisai Co., Ltd. | AI in drug design[125] | Japan | |
| | | 124 | Roche Holding AG | Harnessing the power of AI[126] | Switzerland | Europe |
| | | 125 | Novartis International AG | Ethical and responsible use of AI[127] | Switzerland | |

| | | | | | |
|---|---|---|---|---|---|
| | | 126 | EMS S/A | AI in EMS the future is here[128] | Brazil | South America |
| | | 127 | Sanofi | Sanofi Responsible AI principles[129] | Brazil | |
| | | 128 | CSL Limited | Artificial intelligence at CSL[130] | Australia | Australia |
| | | 129 | Fidson | Integrating tech into healthcare profitable – Firm[131] | Nigeria | Africa |
| 12 | Social Media and Networking/ telecommunications (10) | 130 | Twitter (X) | Synthetic and manipulated media policy[132] | USA | North America |
| | | 131 | LinkedIn Corporation | Linkedin engineering responsible AI[133] | USA | |
| | | 132 | Alibaba Group Holding Limited | Alibaba Cloud Unveils New AI Model to Support Enterprises' Intelligence Transformation[134] | China | Asia |
| | | 133 | WeChat | Wechat AI privacy policy[135] | China | Asia |
| | | 134 | Spotify Technology S.A. | Spotify using AI[136] | Sweden | Europe |
| | | 135 | Skype | Skype Translator AI policy[137] | Luxemburg | |
| | | 136 | MercadoLibre, Inc | Improving user experience and boost sales with AI[138] | Argentina | South America |
| | | 137 | América Móvil | The ITU recognizes the Carlos Slim Foundation and América Móvil for their technological innovation in health care[139] | Mexico | |
| | | 138 | Accel IT | Accel's AI Investments Keep The Focus On Applications And Tooling[140] | Australia | Australia |
| | | 139 | MTN SA | MTN SA introduces Siya as it intensifies its AI Strategy in a bid to boost efficiency and revenue[141] | Nigeria | Africa |
| 13 | Advertising and marketing (10) | 140 | WPP plc | AI regulation is about finding the right balance[142] | USA | North America |
| | | 141 | Omnicom Group Inc. | Omnicom first mover access to AI insights[143] | USA | |
| | | 142 | Dentsu Group Inc. | Dentsu AI innovations[144] | Japan | Asia |
| | | 143 | Hakuhodo DY Holdings Inc. | Interview with the CFO on artificial intelligence[145] | Japan | |
| | | 144 | Publicis Groupe SA | PUBLICIS IS PUTTING AI AT ITS CORE TO BECOME THE INDUSTRY'S FIRST INTELLIGENT SYSTEM[146] | France | Europe |
| | | 145 | Havas Group | AI makes its mark at Havas group[147] | France | |
| | | 146 | Grupo ABC | AI resource guide[148] | Brazil | |

| | | 147 | DPZ&T | Domino's (DPZ) Boosts AI Capabilities With Microsoft Partnership[149] | Brazil | South America |
|---|---|---|---|---|---|---|
| | | 148 | Clemenger Group | Bad News? Send an AI. Good News? Send a Human[150] | Australia | Australia |
| | | 149 | Ogilvy Africa | Creativity, business and society in the age of AI[151] | South Africa | Africa |
| 14 | Legal Tech/ Legal Services/ Intellectual Property Law (10) | 150 | LegalZoom | LegalZoom Launches Doc Assist in Beta, Combining the Power of GenAI and Our Independent Attorney Network[152] | USA | North America |
| | | 151 | Thomson Reuters | AI principles[153] | USA | |
| | | 152 | Zegal | Comprehensive impact of AI[154] | Hong Kong | Asia |
| | | 153 | Cyril Amarchand Mangaldas | Legal Technology & Alternative Legal Services Guideline [155] | India | |
| | | 154 | Rocket Lawyer | AI workplace use policy[156] | UK | Europe |
| | | 155 | Allen & Overy | Artificial Intelligence Use Case at A&O[157] | UK | |
| | | 156 | Demarest Advogados | Good practices using AI[158] | Brazil | South America |
| | | 157 | Lopes Pinto, Nagasse | AI, Data Protection & Privacy 2024 legislation[159] | Argentina | |
| | | 158 | Lawpath | Artificial Intelligence: What is it and Can it Help My Business?[160] | Australia | Australia |
| | | 159 | Webber Wentzel | Webber Wentzel embraces Generative AI as part of its ongoing innovation journey[161] | South Africa | Africa |
| | | 160 | ENS Africa | Responsible AI: embracing generative artificial intelligence technologies- a brief guide for organisations[162] | South Africa | |

**Conflict of interest**: The authors declare that the research was conducted in the absence of any commercial or financial relationships that could be construed as a potential conflict of interest.

**Acknowledgements:** This research is funded by the National Science Foundation (NSF) under grant number 2125858. The authors express their gratitude for the NSF's support, which made this study possible. Furthermore, in accordance with MLA (Modern Language Association) guidelines, we note the use AI-powered tools, such as OpenAI's applications, for assistance in editing and brainstorming.

**Institutional Review Board Statement:** Not applicable.
**Informed Consent Statement:** Not applicable.